\documentclass{PoS}
\usepackage[fleqn]{amsmath}
\usepackage{amssymb}
\usepackage{rsfs}
\usepackage{graphicx}
\usepackage{bm}
\usepackage{cite}
\usepackage{multicol}
\title{
\begin{picture}(0,0)(0,0)%
\put(250,35){\makebox(0,0)[l]
{\textnormal{\normalsize 
CHIBA-EP-232, KEK Preprint 2018-78}}}%
\end{picture}%
Type of dual superconductivity for $SU(2)$ and $SU(3)$ Yang--Mills theories}

\ShortTitle{Type of dual superconductivity}

\author{\speaker{Akihiro Shibata}\\
	Computing Research Center, High Energy Accelerator Research Organization (KEK)\\
	SOKENDAI (The Graduate University for Advanced Studies), Tsukuba 305-0801, Japan\\
        E-mail: \email{Akihiro.shibata@kek.jp}}

\author{Kei-Ichi Kondo\\
	Department of Physics, Graduate School of Science and Engineering, Chiba University\\
	Department of Physics, Graduate School of Science, Chiba University, Chiba 263-8522, Japan\\
	E-mail: \email{kondok@faculty.chiba-u.jp}}

\author{Shogo Nishino\\
	Department of Physics, Graduate School of Science, Chiba University, Chiba 263-8522, Japan\\
	E-mail: \email{shogo.nishino@chiba-u.jp}}

\author{Takaaki Sasago\\
	Department of Physics, Graduate School of Science, Chiba University, Chiba 263-8522, Japan
	}

\author{Seikou Kato\\
	Oyama National College of Technology, Oyama 323-0806, Japan\\
	E-mail: \email{skato@oyama-ct.ac.jp}}

\abstract{We investigate the type of dual superconductivity responsible for quark confinement.
For this purpose, we solve the field equations of the $U(1)$ Abelian--Higgs model to obtain the static vortex solution in the whole range without restricting to the long-distance region.
Then we use the resulting magnetic field of the vortex to fit the gauge-invariant chromoelectric field connecting a pair of quark and antiquark which was measured by numerical simulations for $SU(2)$ and $SU(3)$ Yang--Mills theories on a lattice.
This result improves the accuracy of the fitted value for the Ginzburg--Landau parameter to reconfirm the type I dual superconductivity for quark confinement, which was claimed by preceding works based on an approximate method based on the Clem ansatz.
Moreover, we calculate the Maxwell stress tensor for the fitted model to obtain the distribution of the force around the flux tube.
This suggests that the attractive force acts on the surface perpendicular to the chromoelectric flux tube, in agreement with the type I dual superconductivity.}

\FullConference{XIII Quark Confinement and the Hadron Spectrum - Confinement2018\\
		31 July - 6 August 2018\\
		Maynooth University, Ireland}

\begin{document}
%%%%%%%%%%%%%%%%%%%%%%%%%%%%%%%%%%%%%%%%%%%%%%%%%%%%%%
\section{Introduction}
%%%%%%%%%%%%%%%%%%%%%%%%%%%%%%%%%%%%%%%%%%%%%%%%%%%%%%

From the viewpoint of the dual superconductivity picture, the type of dual superconductor characterizes a property of the vacuum of the Yang--Mills theory or QCD for quark confinement.
In the context of the usual superconductor, in \textbf{type I\hspace{-.1em}I} the repulsive force works among the vortices, while in \textbf{type I} the attractive force works among them.
The boundary of the type I and type I\hspace{-.1em}I is called the \textbf{Bogomol'nyi--Prasad--Sommerfield (BPS) limit} and no forces work among the vortices.

The type of dual superconductor has been investigated for a long time by fitting the chromoelectric flux obtained by lattice simulations to the magnetic field of the ANO vortex.
The preceding studies \cite{type2} done in 1990's concluded that the vacuum of the Yang--Mills theory is of type I\hspace{-.1em}I or the border of type I and type I\hspace{-.1em}I as a dual superconductor.
The improved studies \cite{koma} conclude that the vacuum of the Yang--Mills theory is weakly of type I.
In these studies, however, the fitting range was restricted to a long-distance region from the flux tube.
Recent studies \cite{Kato-Kondo-Shibata, Shibata-Kondo-Kato-Shinohara, type1} show that the vacuum of QCD is the type I dual superconductor.
In these papers, they modify the preceding method by adopting the \textbf{Clem ansatz} \cite{Clem} for incorporating the short distance behavior of the flux tube.
The Clem ansatz assumes the behavior of the complex scalar field (as the order parameter of a condensation of the Cooper pairs), which means that it still uses an approximation.
In this work, we shall fit the chromoelectric flux tube to the magnetic field of the ANO vortex in the $U(1)$ Abelian--Higgs model without any approximations to examine the type of dual superconductor.

In addition, in order to estimate the interaction between the flux tubes, we consider the Maxwell stress carried by a single vortex configuration.
Recently, the Maxwell stress distribution around the quark-antiquark pair was directly observed on a lattice via the gradient flow method \cite{EMT}.
Our results should be compared with their observation.
In order to do this, we shall consider the energy-momentum tensor of a single vortex solution \cite{EMT2} and obtain the Maxwell stress distribution around the vortex with the fitted values of the Ginzburg--Landau parameter.

%%%%%%%%%%%%%%%%%%%%%%%%%%%%%%%%%%%%%%%%%%%%%%%%%%%%%%%%%
\section{Operator on a lattice to measure the flux tube}
%%%%%%%%%%%%%%%%%%%%%%%%%%%%%%%%%%%%%%%%%%%%%%%%%%%%%%%%%

\begin{figure}[t]
\centering
\includegraphics[width=0.5\textwidth]{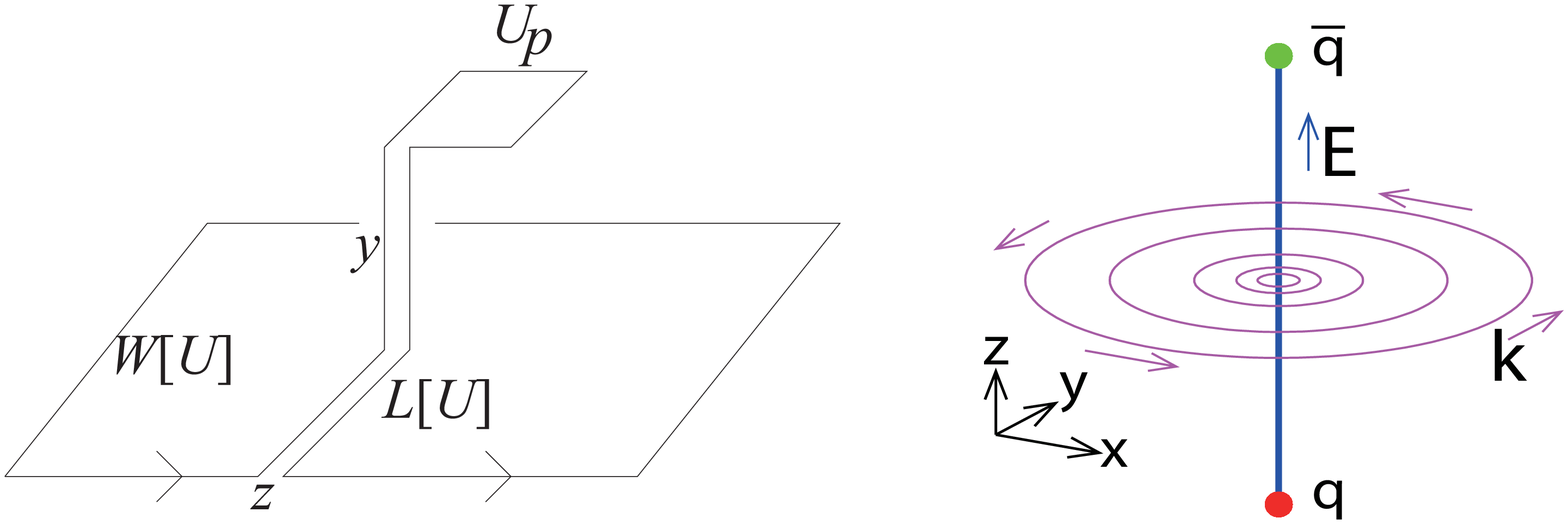}
\caption{(Left panel) The setup of the operator $W [V] L_{V} V_{P} L_{V}^{\dagger}$ in (\ref{lattice_operator}).
$z$ is a position of the Schwinger line $L$, and $y$ is the distance from the Wilson loop $W [V]$ to the plaquette $V_{P}$.
(Right panel) The relation among the chromoelectric field $\bm{E}$, the induced magnetic current $\bm{k}$, and the quark-antiquark pair $q \bar{q}$. }
\label{lattice_result1}
\end{figure}
\begin{figure}[t]
\centering
\includegraphics[width=0.8\textwidth]{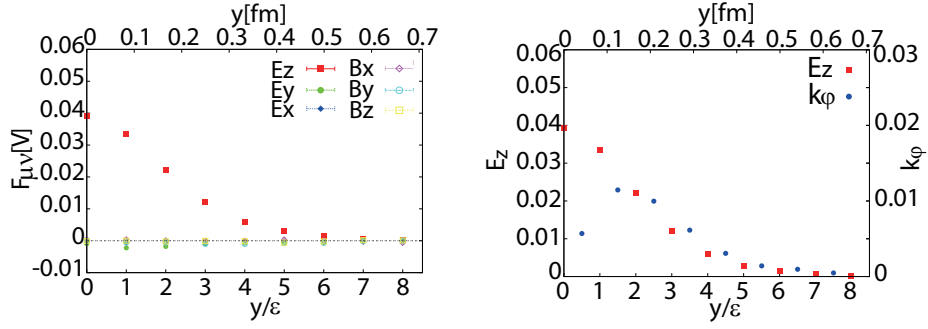}
\caption{(Left panel) The gauge-invariant chromofields measured by using the operator $\rho_{V}$ in (\ref{lattice_operator}) at the midpoint of the $q \bar{q}$ pair ($z = 4$) for the $8 \times 8$ Wilson loop on the $24^{4}$ lattice with the lattice spacing $\epsilon = 0.08320 \  \mathrm{fm}$ at $\beta = 2.5$.
(Right panel) The induced magnetic current $k_{\mu}$ obtained by (\ref{current_lattice}) using the chromofield $F_{\mu \nu} [V]$ for the restricted field $V$.}
\label{lattice_result}
\end{figure}

We have exploited the gauge-invariant operator of Di Giacomo et al.\cite{Giacomo} to measure  chromoelectric and chromomagnetic fields:
% in a gauge-invariant way: (See the above Figure.)
\begin{align}
\rho_{U} := & \frac{\left\langle \mathrm{tr} \left( W [U] L_{U} U_{P} L_{U}^{\dagger} \right) \right\rangle}{\langle \mathrm{tr} ( W [U] ) \rangle} 
- \frac{1}{\mathrm{tr}( \bm{1})} \frac{\langle \mathrm{tr} ( U_{P} ) \mathrm{tr} ( W [U] ) \rangle}{\langle \mathrm{tr} ( W [U] ) \rangle}
,
\end{align}
which is shown in the top left panel of Figure \ref{lattice_result1}.
In the continuum limit $\epsilon \to 0$, $\rho_{U}$ reduces  to
\begin{align}
\rho_{U} = & i g \epsilon^{2} \frac{\left\langle \mathrm{tr} ( \mathscr{F}_{\mu \nu} [\mathscr{A}] L_{U}^{\dagger} W [U] L_{U} ) \right\rangle}{\langle \mathrm{tr} ( W [U] ) \rangle} + \mathcal{O} (\epsilon^{4} ) .
%\nonumber\\
%\simeq & g \epsilon^{2} \langle \mathscr{F}_{\mu \nu} [\mathscr{A}] \rangle_{q \bar{q}} \ \ \ \ ??
\end{align}
This was identified with the chromofield strength generated by a pair of quark and antiquark,  $\rho_{U} \simeq   g \epsilon^{2} \langle \mathscr{F}_{\mu \nu} [\mathscr{A}] \rangle_{q \bar{q}}$.

%We can raise a question whether the gauge-invariant $\rho_{U}$ can extract the color-singlet component of the chromoelectric flux correctly.
%For the gauge group $SU(2)$, indeed, we find 
%\begin{align}
% {\rm tr}[\mathscr{F}_{\mu \nu} L^{\dagger} W L] 
%= {\rm tr}[\mathscr{F}_{\mu \nu}^{A} T_{A} \left( a_{0} \bm{1} +2 i a_{B} T_{B} \right)]
%=  i \mathscr{F}_{\mu \nu}^{A} a_{A},
%\end{align}
%where $\mathscr{F}_{\mu \nu} \in su (2)$ and $L^{\dagger} W L \in SU(2)$ with $a_{0}^2+a_{A}a_{A}=1$.
%
In this paper, we deploy the same operator for the restricted field, which was used  to show the restricted field dominance for the string tension in \cite{Kato-Kondo-Shibata, Shibata-Kondo-Kato-Shinohara}.
We replace the full link variable $U$  by the restricted variable $V$ to define
\begin{align}
\rho_{V} := & \frac{\left\langle \mathrm{tr} \left( W [V] L_{V} V_{P} L_{V}^{\dagger} \right) \right\rangle}{\langle \mathrm{tr} ( W [V] ) \rangle} - \frac{1}{\mathrm{tr} (\bm{1})} \frac{\langle \mathrm{tr} ( V_{P} ) \mathrm{tr} ( W [V] ) \rangle}{\langle \mathrm{tr} ( W [V] ) \rangle}
.
\label{lattice_operator}
\end{align}
It should be noticed that we can define the magnetic current $k_{\mu}$ induced by the chromofield $F_{\mu \nu} [V]$ as
\begin{equation}
k_{\mu} := \frac{1}{2} \epsilon_{\mu \nu \rho \sigma} \nabla_{\nu} F_{ \rho \sigma} [V]
,
\label{current_lattice}
\end{equation}
with the lattice derivative $\nabla_{\nu}$ so that the conservation law $\nabla_{\mu} k_{ \mu} = 0$ holds  \cite{Kato-Kondo-Shibata, Shibata-Kondo-Kato-Shinohara}.
Figure \ref{lattice_result} shows the result of measurement for the $SU(2)$ case \cite{Kato-Kondo-Shibata}.
\section{Fitting method and results}
%%%%%%%%%%%%%%%%%%%%%%%%%%%%%%%%%%%%%%%%%%%%%%%%%%%%%%%%

First of all, we give a brief review of the $U(1)$ Abelian--Higgs model, whose Lagrangian density is given by
\begin{equation}
\mathscr{L} = - \frac{1}{4} F_{\mu \nu} F^{\mu \nu} + \left( D_{\mu} \phi \right)^{\ast} D^{\mu} \phi - \frac{\lambda^{2}}{2} \left( \phi^{\ast} \phi - v^{2} \right)^{2}
,
\label{L_AH}
\end{equation}
where $\lambda$ is the scalar coupling constant and $v$ is the value of the magnitude $|\phi (x)|$ of the complex scalar field $\phi (x)$ in the vacuum.
The asterisk $({}^{\ast})$ denotes the complex conjugation.
The field strength $F_{\mu \nu}$ of the $U(1)$ gauge field $A_{\mu}$ and the covariant derivative $D_{\mu} \phi$ of the scalar field $\phi$ are defined by
\begin{align}
F_{\mu \nu} (x) :=  \partial_{\mu} A_{\nu} (x) - \partial_{\nu} A_{\mu} (x) , \ \ \ 
D_{\mu} \phi (x)  :=  \partial_{\mu} \phi (x) - i q A_{\mu} (x) \phi (x) 
,
\end{align}
where $q$ is the charge of the scalar field $\phi (x)$.
The Euler--Lagrange equations are given by
\begin{align}
\partial^{\mu} F_{\mu \nu} = j_{\nu} , \ \ \ j_{\nu} := i q \bigl[ \phi \left( D_{\nu} \phi \right)^{\ast} - \left( D_{\nu} \phi \right) \phi^{\ast} \bigr] , \ \ \  
D^{\mu} D_{\mu} \phi =  \lambda^{2} \left( v^{2} - \phi^{\ast} \phi \right) \phi
.
\end{align}
In order to obtain the vortex solution, we adopt a static and axisymmetric ansatz:
\begin{equation}
A_{0} (x) = 0 , \ \ \ 
\bm{A} (x) = \frac{n}{q \rho} a (\rho) \bm{e}_{\varphi} , \ \ \ 
\phi (x) = v f (\rho) e^{i n \varphi}
,
\label{ANO_ansatz}
\end{equation}
where we have used the cylindrical coordinate system $(\rho , \varphi , z)$ for the spatial coordinates and $n$ is the winding number.
Notice that the magnetic field $\bm{B} $ can be computed by
\begin{equation}
\bm{B} (x) = \nabla \times \bm{A} (x) = \frac{n}{q \rho} \frac{d a(\rho)}{d \rho} \bm{e}_{z} = : b (\rho) \bm{e}_{z}
.
\end{equation}
We introduce the dimensionless variable $R$ by $R = q v \rho$ and then the functions are reparametrized by $f (\rho) = f (R), \  a (\rho) = a (R) , \  b (\rho) = q^{2} v^{2} b (R) ,  \ j_{\varphi} (\rho) = q^{3} v^{3} j (R)$.
Under this ansatz, the field equations are cast into
\begin{align}
& b^{\prime} (R) +  j (R) = 0 , 
\label{ANO_eq_b} \\
& n a^{\prime} (R) = R b (R) , 
\label{ANO_eq_a} \\
& j (R) = \frac{2n}{R} \left( 1 - a (R) \right) f^{2} (R) , 
\label{ANO_eq_j} \\
& f^{\prime \prime} (R) + \frac{1}{R} f^{\prime} (R) - \frac{n^{2}}{R^{2}} \left( 1- a (R) \right)^{2} f (R) + 2 \kappa^{2} (1 - f^{2} (R) ) f (R) = 0 
\label{ANO_eq_f}
,
\end{align}
where $\kappa := \frac{\lambda}{\sqrt{2} q}$ is the Ginzburg--Landau (GL) parameter and the prime (${}^{\prime}$) denotes the derivative with respect to $R$.
We solve these equations numerically under the following boundary conditions:
\begin{align}
f (0) =  0 , \ \ \ b^{\prime} (0) =  0 , \ \ \ j (0) = 0 , \ \ \ 
 f ( \infty) =  1 , \ \ \  a (\infty) = 1 
.
\end{align}

To determine the type of dual superconductivity for $SU(2)$ Yang--Mills theory, we fit the chromoelectric field and induced magnetic current obtained by the lattice simulation \cite{Kato-Kondo-Shibata} (see the right panel of Figure \ref{lattice_result1} and the right panel of Figure \ref{lattice_result}) to the magnetic field and electric current of the $n=1$ ANO vortex.
In what follows, we denote the lattice data and their errors as $(y_{i} , E_{z}^{L} (y_{i}) , \delta E_{z}^{L} (y_{i}))$ for the chromoelectric field and $(y_{j} , k_{\varphi}^{L} (y_{j}) , \delta k_{\varphi}^{L} (y_{j}))$ for the induced magnetic current.
We introduce the regression functions by
\begin{align}
B (\hat{\rho} ; \hat{\eta}, \hat{\tau} , \kappa) :=  \hat{\eta} b (\hat{\tau} \hat{\rho} ; \kappa) , \ \ \ 
J (\hat{\rho} ; \hat{\eta} , \hat{\tau}, \kappa) :=  \hat{\eta} \hat{\tau} j (\hat{\tau} \hat{\rho} ; \kappa)
,
\end{align}
where $\hat{\rho} := \rho / \epsilon$ is a dimensionless variable, $\hat{\eta} = \eta \epsilon^{2}$ and $\hat{\tau} = \tau \epsilon$ are dimensionless constants with the lattice spacing $\epsilon$.
Here, the $\kappa$-dependence of these functions is implicit, since it is determined once we solve the field equations.

\begin{figure}[t]
\centering
\includegraphics[width=0.4\textwidth]{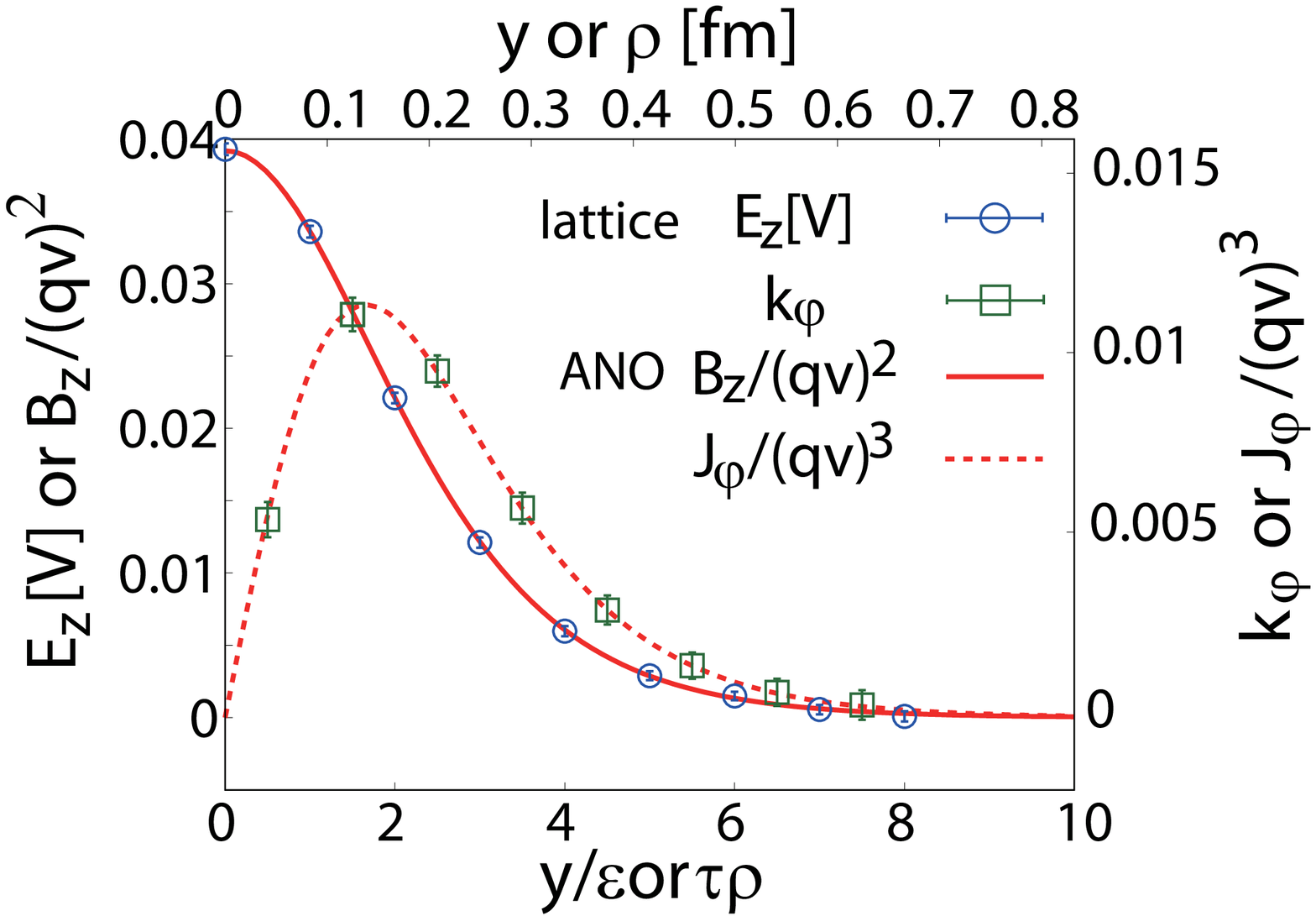} \ \ \ 
\includegraphics[width=0.4\textwidth]{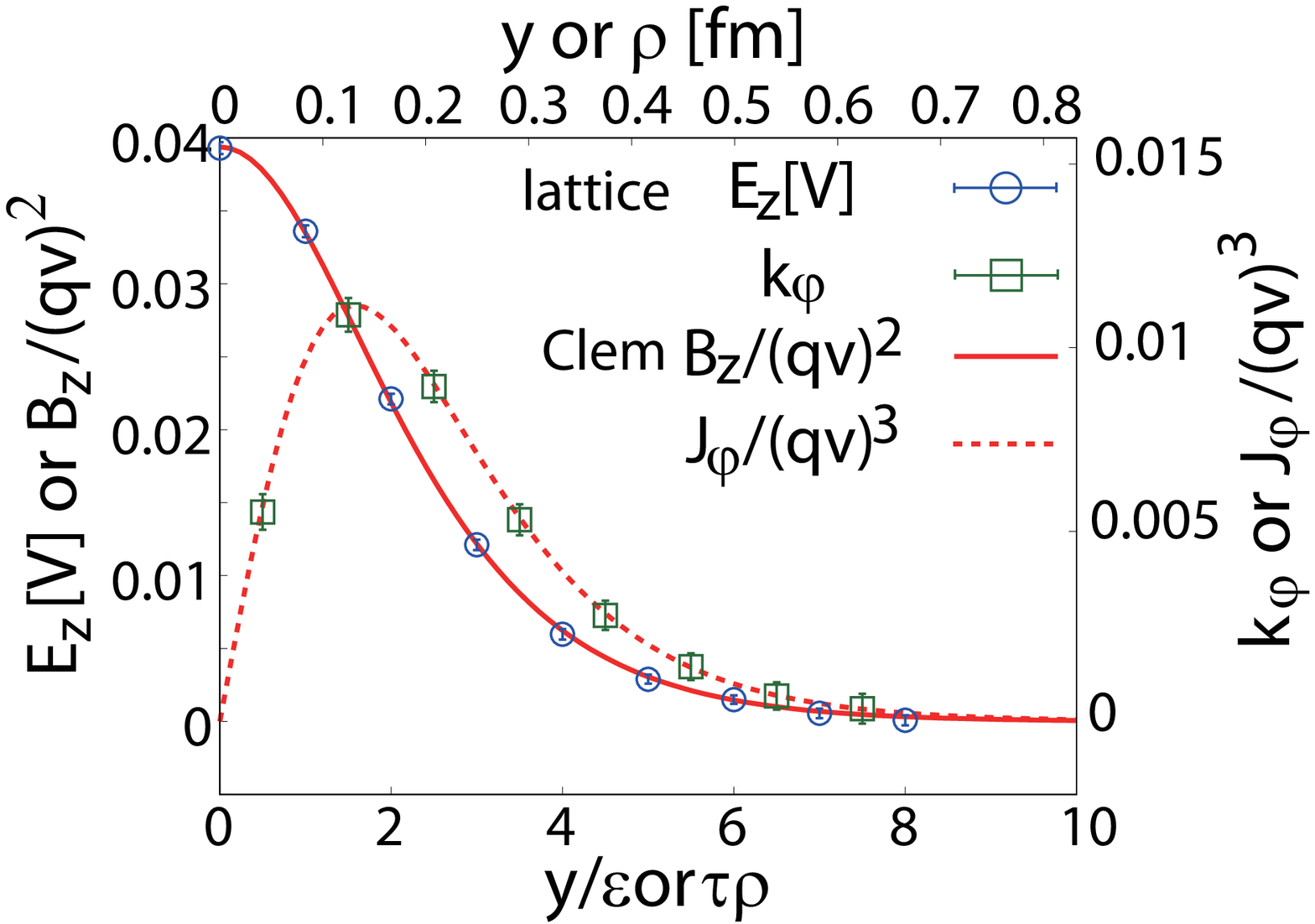} 
\caption{The fitting results: (left panel) the ANO vortex with a unit winding number, (right panel) the approximated method based on the Clem ansatz.}
\label{fit_Figure}
\end{figure}

We adopt the maximal likelihood fitting for the flux and current in (\ref{L_AH}), simultaneously.
The error functions of the regression with the weights are given by
\begin{equation}
\varepsilon_{\rm flux} ( y_{i} ;  \hat{\eta} , \hat{\tau} , \kappa ) = \frac{E_{z}^{L} (y_{i}) - B (y_{i}; \hat{\eta} , \hat{\tau} , \kappa)}{\delta E_{z}^{L} (y_{i}) } , \ \ \ 
\varepsilon_{\rm current} (y_{j} ; \hat{\eta}, \hat{\tau} , \kappa) = \frac{k_{\varphi}^{L} (y_{j}) - J (y_{j} ; \hat{\eta}, \hat{\tau}, \kappa)}{\delta k_{\varphi}^{L} (y_{j})}
.
\label{error}
\end{equation}
When we assume that these error functions follow independent standard normal distributions, the parameters $\hat{\eta} , \hat{\tau}$ and $\kappa$ can be determined by maximizing the log-likelihood function $\ell (\hat{\eta}, \hat{\tau}, \kappa)$:
\begin{equation}
 \ell ( \hat{\eta} , \hat{\tau}  , \kappa ) =- \frac{1}{2} \sum_{i = 1}^{n} \left( \varepsilon_{\rm flux} (y_{i} ; \hat{\eta} , \hat{\tau} , \kappa ) \right)^{2} - \frac{1}{2} \sum_{j = 1}^{m} \left( \varepsilon_{\rm current} (y_{j} ; \hat{\eta}, \hat{\tau} , \kappa ) \right)^{2}
.
\end{equation}
We obtain the result for the ANO vortex with a unit winding number, $n=1$:
\begin{align}
& \hat{\eta}  = 0.0448 \pm 0.0050 , \ \ \ 
\hat{\tau}  = 0.508 \pm 0.032  , \ \ \ 
 \kappa  = 0.565 \pm 0.053 , \nonumber\\
&{\rm MSR}_{\rm flux} = 0.131 , \ \ \ 
{\rm MSR}_{\rm current} = 0.0938 , \ \ \ 
{\rm MSR}_{\rm total} = 0.114 \label{fit_result}
.
\end{align}
where MSR stands for the mean residual sum of squared errors for the regression of (\ref{error}).
The fitting result is shown in the left panel of Figure \ref{fit_Figure}.
This new result shows that the vacuum of $SU(2)$ Yang--Mills theory is of type I.

This result should be compared with result by using the Clem ansatz.
(For more detail, see \cite{preparation}.)
The new result (\ref{fit_result}) gives the larger value of the GL parameter than that in the previous work \cite{Kato-Kondo-Shibata}, $\kappa = 0.38 \pm 0.23$, where only the regression of $E_{z}^{L}$ is taken into account
We also study the improved method based on the Clem ansatz \cite{preparation}, where the fitting for both $E_{z}^{L}$ and $k_{\varphi}^{L}$ is adopted by using the regression function $J (y_{j} ; \alpha , \beta , \kappa ) $ which is replaced by the Clem ansatz.
The fitting result is shown in the right panel of Figure \ref{fit_Figure} and gives the GL parameter:
\begin{align}
&\kappa  = 0.37 \pm 0.20 , \ \ \ 
{\rm MSR}_{\rm flux} = 0.171 , \ \ \ 
{\rm MSR}_{\rm current} = 0.086, \ \ \ 
{\rm MSR}_{\rm total} = 0.135 \label{fit_Clem}
.
\end{align}
The inclusion of $k_{\varphi}$ can improve the accuracy of fitting for the flux.

\section{Type of dual superconductor}
%%%%%%%%%%%%%%%%%%%%%%%%%%%%%%%%%%%%%%%%%%%%%%%%%%%%%%%%%%%%%%%%%%%%%%

\begin{figure}[t]
\centering
\includegraphics[width=1.0\textwidth]{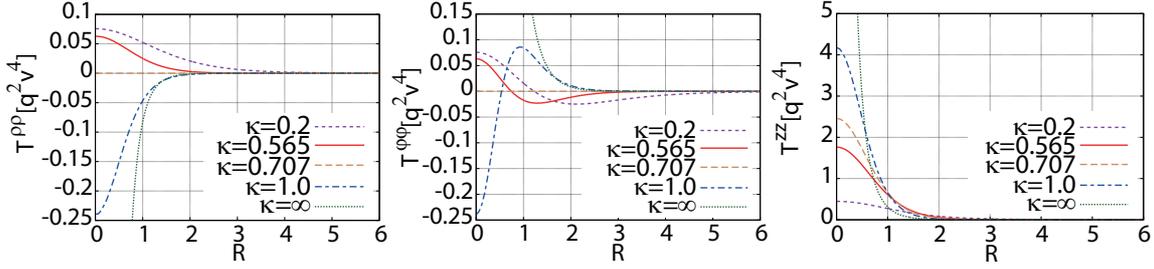}
\caption{The components of the stress tensor $T^{\rho\rho}$ (left panel), $T^{\varphi \varphi}$ (middle panel), and $T^{zz}$ (right panel) as functions of $R$ for the $n=1$ ANO vortex configuration in units of $q^{2} v^{4}$ for $\kappa = \frac{1}{5} , 0.565$ (type I), $\frac{1}{\sqrt{2}}$ (BPS), $1$ (type I\hspace{-.1em}I), and $\infty$ (London limit).
The red solid curves represent the stress tensor for the fitted parameter of the GL parameter $\kappa = 0.565$.}
\label{EMT_ANO_Figure}
\end{figure}

In order to clarify the difference between type I and I\hspace{-.1em}I of dual superconductors, we investigate the Maxwell stress tensor around a vortex according to the proposal \cite{EMT2}.
For this purpose, we obtain the energy-momentum-stress tensor $T^{\mu \nu}$ from the Lagrangian density (\ref{L_AH}) as
\begin{align}
T^{\mu \nu} 
= & \frac{1}{4} g^{\mu \nu} F_{\rho \sigma} F^{\rho \sigma} - F^{\mu \rho} F^{\nu}{}_{\rho} + \left( D^{\mu} \phi \right) \left( D^{\nu} \phi \right)^{\ast} + \left( D^{\mu} \phi \right)^{\ast} \left( D^{\nu} \phi \right) \nonumber\\
& - g^{\mu \nu} \left( D_{\rho} \phi \right) \left( D^{\rho} \phi \right)^{\ast} + \frac{\lambda^{2}}{2} g^{\mu \nu} \left( v^{2} - \phi^{\ast} \phi \right)^{2}
.
\label{EMT_ANO}
\end{align}
Notice that this energy-momentum tensor is symmetric, i.e., $T^{\mu \nu} = T^{\nu \mu}$.
Under the ansatz (\ref{ANO_ansatz}), the components of $T^{\mu \nu}$ are written into
\begin{align}
T^{\rho\rho} = & q^{2} v^{4} \biggl[\frac{1}{2} b^{2} (R) +f^{\prime 2} (R) - \frac{n^{2}}{R^{2}} \left( 1 - a (R) \right)^{2} f^{2} (R) - \kappa^{2} \left( 1 - f^{2} (R) \right)^{2} \biggr] , \\
T^{\varphi\varphi} = & q^{2} v^{4} \biggl[ \frac{1}{2} b^{2} (R) - f^{\prime 2} (R) + \frac{n^{2}}{R^{2}} \left( 1 - a (R) \right)^{2} f^{2} (R) - \kappa^{2} \left( 1 - f^{2} (R) \right)^{2} \biggr] , \\
T^{zz} = & q^{2} v^{4} \biggl[ \frac{1}{2} b^{2} (R) + f^{\prime 2} (R) + \frac{n^{2}}{R^{2}} \left( 1 - a (R) \right)^{2} f^{2} (R) + \kappa^{2} \left( 1 - f^{2} (R) \right)^{2} \biggr] = - T^{00} 
,
\end{align}
and the off-diagonal components vanish.
Figure \ref{EMT_ANO_Figure} shows $T^{\rho \rho}, T^{\varphi \varphi}$ and $T^{zz}$ for various GL parameter $\kappa$ with a unit winding number.
Here, we change the signature of $T^{j k}$ defined in (\ref{EMT_ANO}) by using the ambiguity of the overall signature of the Noether current in order to reproduce the conventional Maxwell stress tensor.

%\begin{figure}[t]
%\centering
%\includegraphics[width=0.8\textwidth]{vortexfig/force_xy2.eps}
%\caption{The stress force $\bm{F}^{\perp}$ acting on the surface element parallel to the magnetic field $\bm{B}$ of the ANO vortex for (Left panel) type I ($\kappa = \frac{1}{5}$), (Mid panel) BPS ($\kappa = \frac{1}{\sqrt{2}}$), and (Right panel) type I\hspace{-.1em}I ($\kappa = 1$).
%The red point stands for the vortex.}
%\label{force_xy}
%\end{figure}
%
%Figure \ref{EMT_ANO_Figure} is the plot of the components of the stress tensor $T^{j k}$ for the $n = 1$ ANO vortex configuration with $\kappa = \frac{1}{5} , \frac{1}{\sqrt{2}}, 1$, and $\infty$.
%One finds that the component $T^{\rho \rho}$ is always positive $T^{\rho \rho} (R) > 0$ in type I, while negative $T^{\rho \rho} (R) < 0$ in type I\hspace{-.1em}I.
%The difference of a signature of $T^{\rho \rho}$ is a key feature of the force around the vortex.
%
%
%
%\begin{figure}[t]
%\centering
%\includegraphics[width=0.33\textwidth]{vortexfig/force_xz_type1.eps} \ 
%\includegraphics[width=0.33\textwidth]{vortexfig/force_xz_type2.eps}
%\caption{The  distibution of the magnitude $|\bm{F}^{/\hspace{-.1em}/}|$ of the stress force $\bm{F}^{/\hspace{-.1em}/}$ on the surface element perpendicular to the magnetic field $\bm{B}$ of the ANO vortex for (Left panel) type I ($\kappa = \frac{1}{2}$) and (Right panel) type I\hspace{-.1em}I ($\kappa = 1$).
%The blue arrows represent $\bm{F}^{\perp}$ in Figure \ref{force_xy}.
%The red line stands for the vortex.}
%\label{force_xz}
%\end{figure}

\begin{figure}[t]
\centering
\includegraphics[width=0.65\textwidth]{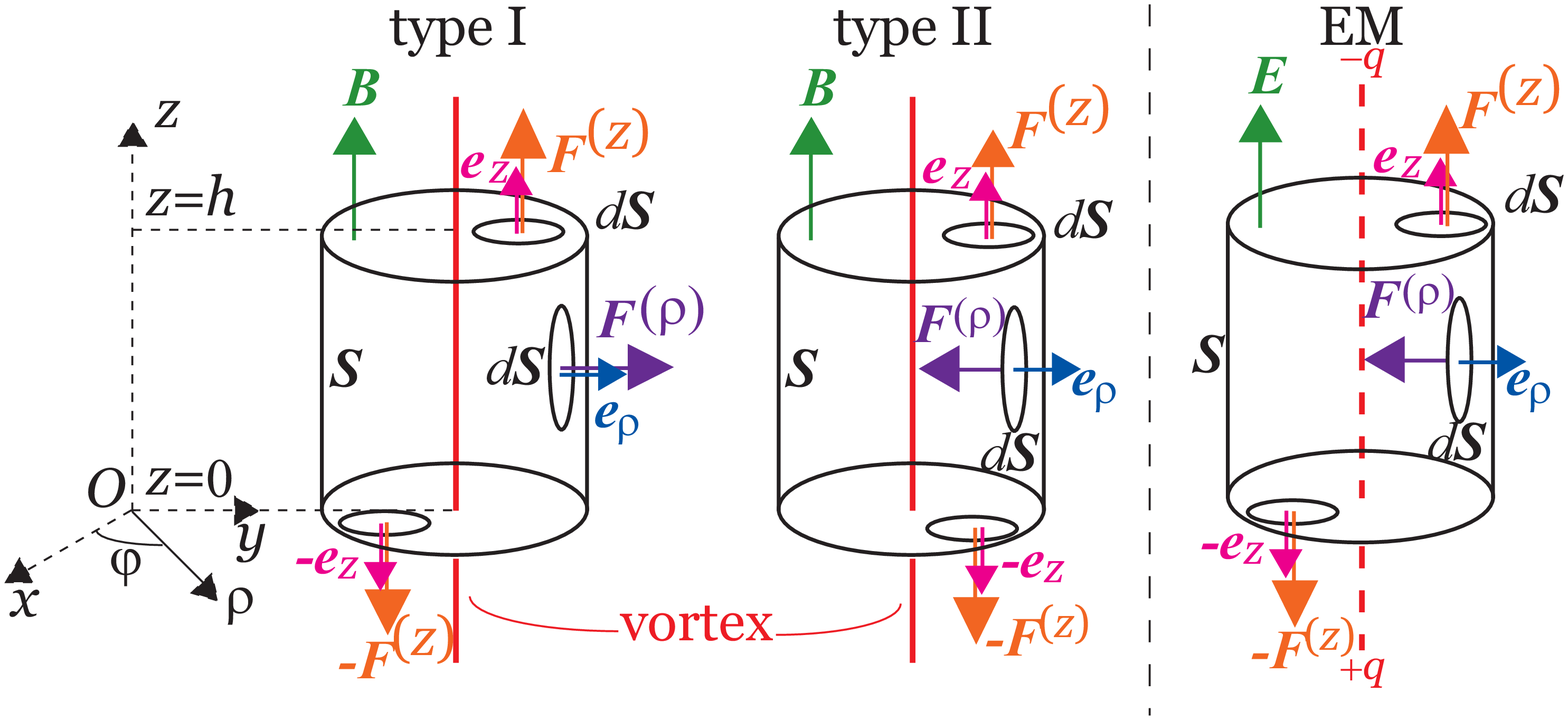}
\caption{(Left and Mid panels) The Maxwell stress force acting on the flux tube originating from the ANO vortex configuration. 
(Right panel) The Maxwell stress force in the electromagnetism.}
\label{force_EM}
\end{figure}

Next, we consider the force acting on the area element of the flux tube.
By using the Maxwell stress tensor, the stress force $\bm{F}$ acting on the infinitesimal area element $d \bm{S}$ is given by
\begin{equation}
\bm{F} = T \cdot d \bm{S} = T \cdot \bm{n} \Delta S
,
\end{equation}
where $\bm{n}$ is a normal vector of the area element $d S$ and $\Delta S$ stands for the area of $d S$.
Figure\ref{force_EM} shows elements of the stress force.
The left and mid panels show the situations for the ANO vortex, while the right panel shows the corresponding situation in the electromagnetism case, where a pair of electric charges $\pm q$ is located at $\mp \infty$ on the $z$-axis, respectively.

If we choose $\bm{n}$ to be equal to the normal vector pointing the $\rho$-direction, i.e., $\bm{n} = \bm{e}_{\rho}$, the corresponding stress force $\bm{F}^{(\rho)}$ reads
\begin{equation}
\bm{F}^{(\rho)} = T^{\rho \rho} \Delta S \bm{e}_{\rho}
.
\end{equation}
We find that $\bm{F}^{(\rho)} \cdot \bm{e}_{\rho} = T^{\rho \rho} \Delta S$ is always positive in type I, while always negative in type I\hspace{-.1em}I. 
Therefore,  $\bm{F}^{(\rho)}$ represents the attractive force for type I, while the repulsive force for type I\hspace{-.1em}I.

The other choice of $\bm{n}$ is to be parallel to the ANO vortex, i.e., $\bm{n} = \bm{e}_{z}$.
The corresponding stress force $\bm{F}^{(z)}$ can be written as
\begin{align}
\bm{F}^{(z)} = T^{z z} \Delta S \bm{e}_{z} , \ \ \ 
\bm{F}^{(z)} \cdot \bm{e}_{z} = T^{zz} \Delta S > 0
.
\end{align}
Figure \ref{force_EM} is a sketch of the Maxwell stress force acting on the flux tube originating from the ANO vortex configuration.

\begin{figure}[t]
\centering
\includegraphics[width=0.5\textwidth]{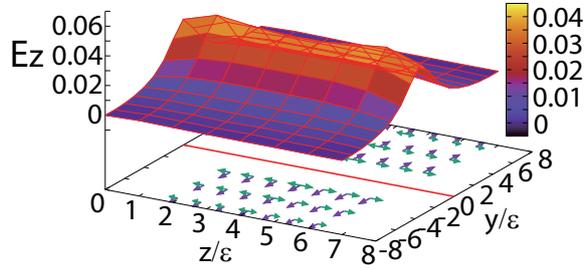}
\caption{The chromoelectric flux obtained in \cite{Kato-Kondo-Shibata} and distribution of the Maxwell stress for the fitted value of the GL parameter $\kappa = 0.564$.
The red line (the thick line in the $z-y$ plane) stands for the ANO vortex.
}
\label{distribution}
\end{figure}

Using the parameters obtained by fitting to the ANO vortex, we can show the distribution of the Maxwell stress around the flux tube, which is shown in Figure \ref{distribution}.
This result indeed supports the type I dual superconductor for quark confinement.

Our analysis on the Maxwell stress tensor around an ANO vortex agrees with the result obtained by the preceding work \cite{EMT2}.

%%%%%%%%%%%%%%%%%%%%%%%%%%%%%%%%%%%%%%%%%%%%%%%%%%%%%%%%%%%
\section{Conclusion}
%%%%%%%%%%%%%%%%%%%%%%%%%%%%%%%%%%%%%%%%%%%%%%%%%%%%%%%%%%%

We investigate the type of dual superconductivity responsible for quark confinement. 
For this purpose, we have solved the field equations of the $U(1)$ Abelian--Higgs model without any approximation in place of Clem ansatz, and have fitted the flux and magnetic current.
We have reconfirmed that the vacuum of the $SU(2)$ Yang--Mills theory is of type I as a dual superconductor with the GL parameter $\kappa=0.565\pm0.053$.
We found that inclusion of regression of the magnetic current is important to improve the accuracy of the fitting as seen from the error of the GL parameter, or the mean of squared residuals.
We also found that the approximated method based on the Clem ansatz is sensitive to the fitting range.
In the new method, on the other hand, the effect of changing the fitting range is negligible.
This fact suggests that our new method gives more reliable results than the previous one.
For more detail, see \cite{preparation}.

Moreover, we have calculated the distribution of the Maxwell stress force around the flux tube for the Abelian--Higgs model with the fitted GL parameter.
It was confirmed that there exists an attractive force among the chromoelectric flux tubes, that is consistent with the type I dual superconductor.

\section*{Acknowledgement}
The authors would like to thank Hideo Suganuma for valuable discussions, especially suggestions on error estimations.
They would like to express sincere thanks to Ryosuke Yanagihara, Takumi Iritani, Masakiyo Kitazawa, and Tetsuo Hatsuda for very helpful and illuminating discussions on the Maxwell stress tensor in the early stage of their investigations, on which a part of the result presented in section V is based.
This work was supported by Grant-in-Aid for Scientific Research, JSPS KAKENHI Grant Number (C) No.15K05042.
S.N. thanks Nakamura Sekizen-kai for a scholarship.

\end{document}